\documentclass[12pt]{article}

\usepackage{amsmath}
\usepackage{graphicx}
\usepackage{overcite}
\usepackage{float}

\begin{document}

\author{Moritz P. Haag and Markus Reiher\footnote{Corresponding author;
        e-mail:markus.reiher@phys.chem.ethz.ch} \\
        \small ETH Zurich, Laboratory of Physical Chemistry, Wolfgang-Pauli-Strasse 10   \\
        \small 8093 Zurich, Switzerland}
\title{Real-time Quantum Chemistry}
\maketitle

\begin{abstract}
Significant progress in the development of efficient and fast algorithms for quantum chemical 
calculations has been made in the past two decades. The main focus has always been the desire to be 
able to treat ever larger molecules or molecular assemblies---especially linear and sub-linear 
scaling techniques are devoted to the accomplishment of this goal. However, as many chemical 
reactions are rather local, they usually involve only a limited number of atoms so that models of 
about two hundred (or even less) atoms embedded in a suitable environment are sufficient to study their
mechanisms. Thus, the system size does not need to be enlarged, but remains 
constant for reactions of this type that can be described by less than two hundred atoms.
The question then arises how fast one can obtain the quantum chemical results. This question 
is not directly answered by linear-scaling techniques. In fact, ideas such as haptic quantum 
chemistry or interactive quantum chemistry require an immediate provision of quantum chemical 
information which demands the calculation of data in ``real time''. In this perspective, we aim at a 
definition of real-time quantum chemistry, explore its realm and eventually discuss applications in 
the field of haptic quantum chemistry. For the latter we elaborate whether a direct approach is 
possible by virtue of real-time quantum chemistry.
\vspace{1cm}
\noindent
\begin{center}\it
Submitted for publication as a perspective article in Int.\ J.\ Quantum Chem.
\end{center}
\vspace{1cm}
\end{abstract}

\newpage

\renewcommand{\baselinestretch}{1.5}
\normalsize

\section{Introduction}

The large number of possible nuclear configurations puts the curse of dimensionality\cite{bellman1957}
on studies of the chemical reactivity of large molecular systems like metalloenzymes or transition metal 
complexes in homogeneous catalysis. To find exactly those configurations which correspond to
the reaction path searched for is very difficult. Although there exist several methods to sample
the configuration space of the nuclear positions efficiently and in an unbiased way (see, e.g., 
Refs.~\citen{huber1994,dellago1998,iannuzzi2003,wales2006landscapes}), a full {\it ab initio} treatment of 
the molecular system renders also these approaches very time demanding or even unfeasible. An ab 
initio treatment is, however, mandatory for the study of chemical reactivity involving the forming 
and breaking of chemical bonds. 

The overwhelming amount of data generated by computational methods calls for new approaches to 
access it. In conventional methods, a change of nuclear coordinates $\{\boldsymbol{R}_I\}$ of a reactive or functional molecular system 
changes the potential energy\cite{wales2006landscapes} 
\begin{align}\label{eq:totenergy}
 E_\textnormal{BO}(\{\boldsymbol{R}_I\}) = E_\textnormal{el}(\{\boldsymbol{R}_I\}) +
 \sum\limits_{I,J>I}^M \frac{Z_I\,Z_J}{\left| \boldsymbol{R}_J - \boldsymbol{R}_I \right|} \,\, ,
\end{align}
i.e., the electronic energy defined in the Born--Oppenheimer (BO) approximation.
Once this energy has been calculated by a computer program,
the results are collected and can be evaluated. However, if the results were 
accessible immediately after applying the configurational change, the perception of the results of 
the calculations would be much more efficient since a real interaction between the scientist and 
the system under study could be established. 

In this perspective article we discuss a point of view on such a challenge that we may 
summarize under the term Real-time Quantum Chemistry. ``Real-time'' in this context means that the 
results of quantum chemical calculations are basically instantaneously available, i.e., on a time 
scale that a specific human sense would interpret as instantaneous. Quantum chemical information 
available in real time then offers a very efficient way to interactively study the reactivity of 
molecular systems.

The structure of this work is as follows. We start with a definition of Real-time Quantum Chemistry 
and discuss its principles. Then, we review and evaluate existing techniques to accelerate quantum 
chemical calculations with respect to their potential usage in Real-time Quantum Chemistry. Finally, we 
discuss the feasibility of a direct approach to Haptic Quantum Chemistry\cite{marti2009,haag2011} 
which implements the ideas of Real-time Quantum Chemistry. In order to illustrate what can already be done 
with currently available quantum chemistry software we present some exploratory calculations.

\section{Principles of Real-time Quantum Chemistry}\label{sec:rtqc}

A theoretical study of chemical reactivity requires the exploration of the potential energy surface.
Hence, first the wave function of the molecular system has to be calculated, from which then the
energy and the gradients can be obtained. Other properties like molecular orbital representations,
the electron density, polarizabilities or partial charges could also be considered, but are 
secondary targets compared to energies and gradients. We thus define

\begin{center}
\fbox{\parbox{\dimexpr \linewidth - 20\fboxrule - 20\fboxsep}{
{\bf Definition}{~ \it Real-time Quantum Chemistry shall denote the very fast calculation of 
the quantum mechanical response of a reactive molecular system in terms of the wave function and the 
corresponding energy and gradient due to a user-driven manipulation of the system's molecular 
structure such that the operator experiences the information flow without any time delay.}
}}
\end{center}

Since finding the optimized wave function and calculating the energy and the gradient is essential
to all reactivity studies, we define their instantaneous calculation as the core objective of 
Real-time Quantum Chemistry. Obtaining additional properties in real time could then be 
considered an extension of (core) Real-time Quantum Chemistry.

It is clear that the fast calculation of the core quantities is central also to {\it ab initio} molecular 
dynamics (AIMD)\cite{marx2009}. Accordingly, the challenges which one faces in the context of 
Real-time Quantum Chemistry can also be found in the area of AIMD, where the fast calculation
of the nuclear gradients (called ionic forces) is of paramount importance. Therefore, many 
techniques developed in AIMD to speed up the calculation of gradients are also important in a 
Real-time Quantum Chemistry framework. Although the forces are required as quickly as possible, AIMD 
simulation results are interpreted {\it after} the simulation has been carried out. In Real-time Quantum Chemistry, 
however, we want to seamlessly merge the calculation and the perception of the calculated 
information.

The seamless integration of complex quantum chemical information allows the interactive exploration 
of chemical reactivity. An approach which utilizes a force-feedback device as an input and an output 
device to transmit the quantum mechanical information to the operator is Haptic Quantum Chemistry 
(HQC) introduced by us in 2009\cite{marti2009,haag2011}. 
A force-feedback device as in HQC is also utilized in the interactive molecular dynamics framework\cite{grayson2003}, which 
implements a classical treatment of the forces in molecular systems.
Another example is the recently presented semi-empirical interactive implementation for optimizing molecular structures of 
hydrocarbons during editing the structures\cite{bosson2011,bosson2012}. 

As it has been noted in Ref.~\citen{bosson2012} a real-time {\it visual} experience requires at 
least ten frames per second (i.e.\ $100\,\textnormal{ms}$ to update the system's state), if a frame
represents a step in the shift of nuclear positions during structure optimization or reaction. 
In Haptic Quantum Chemistry, however, a smooth tactile experience requires an update rate of 1000 
frames per second. Accordingly, the central question for real-time quantum chemical approaches is 
the following: Is it---in principle---possible to perform sufficiently accurate {\it ab initio} quantum 
chemical calculations in such a short time, i.e., in the millisecond range for a reasonably sized 
molecular model of a reactive system?

The issue of accelerating quantum chemical calculations has been treated thoroughly before---mostly 
in the area of linear and sub-linear scaling techniques\cite{goedecker1999,rubensson2011}. However, 
it is important to note that there is a fundamental difference between the problem formulated in the 
question above and the problem of achieving linear scaling. In Real-time Quantum Chemistry, we do 
{\it not} ask for methods which allow us to treat larger and larger systems but rather how fast we 
can perform a calculation for a particular system of constant size and target accuracy. This implies 
that one also has to focus on the prefactors and onsets of quantum chemical methods and not only on
their overall scaling behavior. In other words, the actual execution time for a given system is the 
prime target.

The reason why we focus on reactive systems with a (large but) constant number of atoms is that we 
concentrate on rather local chemical events, which are ubiquitous in chemistry. For example, they 
can be mediated by transition-metal centers as in enzymatic reactions or in homogeneous catalysis. 
Reactions at transition metals are usually restricted to a limited spatial region containing the 
metal center(s), its (their) ligand environment, the incoming reactant, co-factors, other reactants 
that may lead to side reactions and some proper model of the close environment in such a way that models on 
the order of $200$ atoms are sufficient for a meaningful description\cite{siegbahn2011,podewitz2011}. 

\section{Methods to Accelerate Electronic Structure Calculations}\label{sec:methodreview}

Almost all acceleration techniques for electronic structure calculations developed so far aim at a
linear scaling behavior especially for very large molecules ($> 1000$ atoms). A multitude of methods 
has been developed as discussed in many excellent reviews\cite{goedecker1999,rubensson2011,
ochsenfeld2007}. Here, we need to evaluate the existing methods from a different point of view. The 
overall goal is to decide, whether there is a certain lower limit of computation time for mid-sized 
molecular systems ($50 - 200$ atoms), and to identify approaches that are likely to be important for a real-time 
calculation of gradients and energies. 

In view of the fact that single-determinant models like Hartree--Fock theory and, most importantly,
density functional theory (DFT) are likely to be the best candidates for quantitative real-time
reactivity exploration, we first need to consider the solution of the Roothaan--Hall equations. The 
two most important steps in a self-consistent solution of the Roothaan--Hall equations are the
construction of the Fock matrix and the subsequent calculation of the density matrix. Of course the 
size of the basis set chosen is also very important, since it determines the size of the system.

We skip the derivation of the one-particle mean-field Hartree--Fock theory\cite{szabo1996} and of 
Kohn--Sham (KS) density functional theory\cite{hohenberg1964,kohn1965,parr1994} which lead both to an effective one-particle
operator equation. For the sake of simplicity, the following equations are given for spin-restricted
calculations. The generalization to the spin-unrestricted case is straightforward. The Fock operator
$\hat{f}$  for Hartree--Fock and Kohn--Sham models can be written as\cite{rudberg2011,rubensson2011}
\begin{align}
 \hat{f}(\boldsymbol{r}_1) = \hat{h}(\boldsymbol{r}_1) +
\sum\limits_a^{N/2}\left[2\,\hat{J}_a(\boldsymbol{r}_1) - \gamma\,\hat{K}_a(\boldsymbol{r}_1)\right]
+ \lambda\,\hat{v}^\textnormal{xc}(\boldsymbol{r}_1) \,\, ,
\end{align}
where $\gamma$ and $\lambda$ are two parameters that define the electronic structure model. For a 
Hartree--Fock calculation, one chooses $\lambda = 0$ and $\gamma = 1$, whereas for a 'pure' Kohn--Sham 
calculation $\gamma = 0$ and $\lambda = 1$ with a non-vanishing $v^\textnormal{xc}$ hold. With 
$\lambda = 1$ and $\gamma \in [0,1]$ hybrid approaches are described.

The introduction of a finite basis set $\{\phi_\nu\}$, 
\begin{align}\label{eq:finitebasis}
 \psi_i(\boldsymbol{r}) = \sum\limits_\nu^K C_{\nu i}\,\phi_\nu(\boldsymbol{r})
\end{align}
is the most convenient way to solve the spatial integro-differential self-consistent-field (SCF) 
equations that result when $\hat{f}$ is operating on an orbital $\psi_i$. This yields the well-known 
Roothaan--Hall equations, 
\begin{align}\label{eq:roothaan}
 \boldsymbol{F}\,\boldsymbol{C} = \boldsymbol{S}\,\boldsymbol{C}\,\boldsymbol{\epsilon} \,\, ,
\end{align}
where $\boldsymbol{C}$ is the matrix of the expansion coefficients defined in 
Eq.~(\ref{eq:finitebasis}), $\boldsymbol{S}$ is the overlap matrix and $\boldsymbol{\epsilon}$ is the
matrix of the orbital energies $\epsilon_i$ of the orbitals $\psi_i$. The elements of the Fock matrix 
$\boldsymbol{F}$ in the chosen basis $\left\{\phi_\nu\right\}$ are given by
\begin{align}
 F_{\mu\nu} = H^\textnormal{core}_{\mu\nu} + J_{\mu\nu} + \gamma\,K_{\mu\nu} + \lambda\,
 V^\textnormal{xc}_{\mu\nu} \,\, .
\end{align}
Here, $\boldsymbol{H}^\textnormal{core} = \boldsymbol{T} + \boldsymbol{V}^\textnormal{eN}$ is the
matrix representation of the one-electron operator, $\boldsymbol{J}$ is the two-electron Coulomb 
matrix operator, $\boldsymbol{K}$ is the two-electron exchange matrix operator and 
$\boldsymbol{V}^\textnormal{xc}$ is the two-electron exchange--correlation matrix operator. For the 
calculation of the matrices $\boldsymbol{J}$ and $\boldsymbol{K}$ the two-electron repulsion 
integrals are contracted with the elements of the density matrix $\boldsymbol{P}$ which is for real 
orbitals in closed-shell systems defined as
\begin{align}
 P_{\mu\nu} = 2 \sum\limits_a^{N/2} C_{\mu a} C_{\nu a} \,\, ,
\end{align}
and thus the reason for the self-consistent iterative solution of the Roothaan--Hall equation.

The elements of the matrix $\boldsymbol{V}^\textnormal{xc}$ are derived from the 
exchange--correlation density functional $\mathcal{F}$ which can be a functional of the electron 
density $\rho$ in the local density approximation, of the density and the gradient of the density 
$\nabla\rho$ in the generalized gradient approximation (GGA), or of the density, the gradient of the 
density, and the kinetic energy density $\tau$ in meta-GGA functionals. Since the matrix elements cannot be 
evaluated analytically the integration is usually performed on a mesh of grid points constructed from merged atomic
grids
\begin{align}
 \left\langle \phi_\mu | \hat{v}^\textnormal{xc} | \phi_\nu  \right\rangle =
 \sum\limits_I^{M} \sum\limits_a^{N^\textnormal{grid}_I} \mathcal{P}_I w_a
 \frac{\partial\mathcal{F}^\textnormal{xc}(\boldsymbol{r}_a)}{\partial \rho(\boldsymbol{r}_a)}
 \phi_\mu(\boldsymbol{r}_a) \,\phi_\nu(\boldsymbol{r}_a) \,\, ,
\end{align}
where $N^\textnormal{grid}_I$ is the number of points $\{\boldsymbol{r}_a\}$ in the grid of atom $I$
and $\{w_a\}$ are the corresponding weights. $\mathcal{P}_I$ is the function that splits the 
molecular grid into atomic sub-grids. The  electron density $\rho$ at a grid-point 
$\boldsymbol{r}_a$ is given by
\begin{align}
 \rho(\boldsymbol{r}_a) = \sum\limits_{\mu\nu} P_{\mu\nu} \phi_\mu(\boldsymbol{r}_a)\,
 \phi_\nu(\boldsymbol{r}_a) \,\, .
\end{align}

Eq.~(\ref{eq:roothaan}) has to be solved iteratively, since $\boldsymbol{F}$ depends on the elements
of $\boldsymbol{C}$. After reaching self consistency the total energy of the system is calculated 
from the converged density matrix elements and the Fock matrix elements,
\begin{align}
 E_\textnormal{BO} &= \sum\limits_{\mu\nu} P_{\nu\mu} H^\textnormal{core}_{\mu\nu} + \frac{1}{2} 
 \sum\limits_{\mu\nu} P_{\nu\mu} \left[ J_{\mu\nu} + \gamma K_{\mu\nu} \right] + 
\sum\limits_{\mu\nu} P_{\nu\mu} V^\textnormal{xc}_{\mu\nu} + V^\textnormal{NN} \,\, ,
\end{align}
where $V^{\rm NN}$ represents the Coulombic pair interaction of all atomic nuclei. From this equation 
also the gradient $\nabla_I$ with respect to the nuclear coordinates $\{\boldsymbol{R}_I\}$ can be derived. Within
a finite basis of, e.g., Cartesian Gaussian functions one obtains after a few rearrangements the
expression\cite{pulay1987,szabo1996}
\begin{align}\label{eq:derivs}
 \nabla_I E_\textnormal{BO} = &\sum\limits_{\mu\nu} P_{\mu\nu} \left( \nabla_I 
 H_{\mu\nu}^\textnormal{core} \right) + \frac{1}{2} \sum\limits_{\mu\nu\lambda\sigma} 
 P_{\mu\nu} P_{\lambda\sigma} \left[ \nabla_I (\mu\nu|\sigma\lambda) - \gamma \frac{1}{2} \nabla_I 
 (\mu\lambda|\sigma\nu) \right] \nonumber \\
 &+ \sum\limits_{\mu\nu} P_{\mu\nu} \left( \nabla_I V^\textnormal{xc}_{\mu\nu} \right) 
 - \sum\limits_{\mu\nu} Q_{\mu\nu} \left( \nabla_I S_{\mu\nu} \right) + \nabla_I V^\textnormal{NN}
\end{align}
It includes the Pulay forces\cite{pulay1987} that are due to the origin dependence of atom-centered basis 
functions. In case of plane-wave basis functions these forces vanish and the expression can be 
simplified. For a more compact notation an energy weighted density matrix $\boldsymbol{Q}$,
\begin{align}
 Q_{\nu\mu} = 2 \sum\limits_a^{N/2} \epsilon_a C_{\mu a} C_{\nu a} \,\, ,
\end{align}
has been introduced. The derivatives of the exchange--correlation matrix elements $\partial 
V_{\mu\nu}^\textnormal{xc} / \partial \boldsymbol{R}_I$ depend on the specific nature of the 
exchange--correlation potential $v^\textnormal{xc}$. General formulations of the energy gradient can
be found in Refs.~\citen{satoko1984,versluis1988, fournier1989, delley1991} for the local (spin)
density approximation and in Ref.~\citen{pople1992} for gradient-corrected functionals.

\subsection{Basis Sets}\label{sec:basisset}

Very important for any acceleration technique is the choice of the basis set in which the Kohn--Sham
or Hartree--Fock orbitals are expanded. The larger the basis set the more accurate the 
calculations are, but also the more time they consume. Reducing their size is, therefore, a very 
seductive means to reduce computation time. In addition certain basis sets speed-up certain parts of 
the Fock matrix calculation but may have disadvantages in other parts. 

The two most widely employed explicit basis sets are linear combinations of Gauss-type atomic 
orbitals and plane waves. Plane waves are especially suited for the calculation of periodic and 
homogeneous systems whereas the Gauss-type orbitals are usually employed in molecular systems. 
Density matrices in Gaussian basis sets are therefore mostly sparse, i.e.\ band diagonal 
and thus promote linear-scaling techniques.

Plane-wave basis sets are completely delocalized in the direct space but are very localized in
reciprocal space, which is why they are usually applied in solid state physics. In molecular
systems, however, a huge number of plane waves would be needed to obtain the same accuracy as with
localized Gauss-type orbitals. On the other hand, some of the integrals can be calculated very 
fast by Fast Fourier Transforms (FFT) within a plane-wave basis set. Also the calculation of the 
forces on the nuclei is computationally less expensive since the basis functions are not position 
dependent. To increase the accuracy and limit the number of basis functions in molecular systems, 
hybrid codes combine Gaussians and plane waves \cite{lippert1997,lippert1999,vandevondele2005}. 
These developments allow for the calculation of molecules with up to a million 
atoms\cite{vandevondele2012}.

Pseudopotentials\cite{schwerdtfeger2011,dolg2012} can be employed for both types of basis sets in 
order to reduce the number basis functions. How many electrons are treated implicitly by the 
pseudopotential can, of course, be varied and determines the accuracy achieved. In plane-wave calculations 
pseudopotentials are crucial to avoid the description of the nodal structure of orbitals so that the 
number of plane-waves can be limited to a reasonably small value. Pseudopotentials are also employed for 
heavy elements to account for relativistic effects\cite{reih09}. 

A less common alternative to the classes of basis sets described above are the so-called wavelets\cite{cho1993,han1999,genovese2008},
which aim at combining the best of both worlds. They are 
localized in both direct and reciprocal space and the integrals can be calculated with very fast 
methods similar to Fast Fourier Transforms. 

It is obvious, that for Real-time Quantum Chemistry the number of basis functions needs to be as 
small as possible. Since high accuracy does not need to be the main goal, comparatively small basis 
sets can be employed. For the same reason pseudopotentials are beneficial. Since plane waves and 
basis sets of plane waves mixed with Gaussians have successfully been applied in AIMD calculations,
they are also useful for Real-time Quantum Chemistry. 

\subsection{Fock Matrix Calculation}\label{sec:fockmatrix}

The calculation of the Fock matrix can be divided into two different parts. The calculation of the 
elements of the one-electron and two-electron matrix operators. 
The calculation of the exchange--correlation matrix elements occurring in KS-DFT calculations will 
be discussed separately. In almost all {\it ab initio} electronic structure calculations the Fock 
matrix construction is the most time consuming part, because of the evaluation of all integrals. 
Therefore, many sophisticated algorithms have been devised to speed up their evaluation.

The calculation of the one-electron Hamiltonian matrix requires the evaluation of $\mathcal{O}(K^2)$ matrix elements,
where $K$ denotes the number of basis functions. The matrix elements consist of two
terms, the kinetic energy term and the electron--nuclei interaction part. If localized basis 
functions are employed, the number of integrals for the kinetic term will increase only linearly
with system size\cite{rubensson2011}. For the Coulomb interaction between the electrons and the 
nuclei, multipole expansions can be applied to reduce the number of integrals and achieve linear 
scaling\cite{rubensson2011}. Although the evaluation of these matrix elements has a quadratic 
scaling, their contribution to the overall execution time is for mid-sized molecular systems very small and hence 
no problem for Real-time Quantum Chemistry. However, to even speed up this part, 
program parallelization can be employed efficiently as the integrals can be evaluated independently
from one another.

The two-electron repulsion integrals (ERI) in the calculation of the Coulomb- and the exchange 
matrix elements are in principle four-index quantities and thus their evaluation formally requires
$\mathcal{O}(K^4)$ operations. The first and most effective way to reduce their
computation time is to discard all elements whose contribution is below a certain threshold.
The most widely employed technique for such an integral screening is based on 
Schwarz-inequality integral estimates.\cite{almloef1982,haeser1989,cremer1986} Another approach to screen 
for negligible matrix elements are the multipole-based integral estimates.\cite{lambrecht2005} They 
consider in addition the $1/R$ decay behavior between two charge distributions. In non-direct SCF 
calculations the integrals are pre-calculated and integral screening can only be done at the level 
of the integrals. In direct SCF methods also the density matrix elements with which the 
integrals are contracted can be taken into account for screening. Therefore, also large integrals 
can be neglected, if they have only a small weight assigned by the density matrix elements. 

The calculation of the surviving integrals can then be accelerated. The first general approach
is to fit the densities occurring in the Coulomb and exchange integrals with auxiliary basis sets, 
thus reducing the four index ERIs to two index quantities. This approach to fit the densities 
accelerates the evaluation of the Coulomb matrix elements by an order of magnitude. To obtain the 
auxiliary basis sets two different methods are employed. One is to determine them by a 
fitting procedure\cite{whitten1973,baerends1973,dunlap1977,dunlap1979}, which is performed before 
the actual electronic structure calculation for each atom type and basis set and results in 
additional libraries of auxiliary basis sets. This density-fitting approach is also known as the resolution of the 
identity (RI)\cite{vahtras1993,eichkorn1995,skylaris2000,weigend2002}. Another method is to employ a 
Cholesky decomposition (CD) algorithm to determine the auxiliary basis set for each calculation 
separately. \cite{beebe1977,koch2003,bostrom2009,aquilante2007,aquilante2011} CD methods are 
computationally more demanding, but generate an auxiliary basis set, which is ``the best'' for a 
given basis and they do not depend on pre-fitted parameters. Therefore, the error introduced by the 
technique is controllable and no extra auxiliary basis-set libraries are needed. 

The calculation of Coulomb matrix elements can also be made more efficient by employing hierarchical
multipole methods like the fast multipole method\cite{greengard1987,white1994,strain1996} or the 
quantum chemical tree code\cite{challacombe1997}. Here, the problem of calculating the Coulomb 
interaction between many electrons is approximated by a truncated multipole expansion. A combination 
of a multipole expansion method and auxiliary basis sets can be applied to reduce the computation 
time even further\cite{sierka2003}.

For the exchange matrix elements special methods exist to obtain a linear-scaling behavior. The RI
technique, which is very efficient for the Coulomb matrix elements, can also be applied for the 
exchange matrix elements not with the same efficiency though\cite{weigend2002}. In addition 
several methods ave been developed specifically for the exchange matrix elements. Examples are the 
$\mathcal{O}(N)$-Exchange method\cite{schwegler1996} (where $N$ denotes the number of basis 
functions and is identical to our $K$), the local K algorithm\cite{aquilante2007b}, the LinK 
method\cite{ochsenfeld1998} or the auxiliary density matrix methods for Hartree--Fock-type exchange\cite{guidon2010}. 

In KS-DFT calculations the Fock matrix $\boldsymbol{F}$ contains additional matrix elements 
from the exchange--correlation functional. The matrix elements are evaluated by numerical 
integration of the functional derivative on a grid. The computational cost depends, of course, on the size
of the grid. The overall molecular grid is constructed from atomic grids, which are 
merged to obtain the molecular grid. A very common choice is the Becke atomic partitioning 
scheme\cite{becke1988b}. Since the calculation for each atomic grid can be done independently the 
overall scaling behavior can be made linear.\cite{vanwuellen1994,perezjorda1995,stratmann1996,
burow2011} The exchange--correlation matrix elements are usually considered as requiring a negligible 
amount of computation time compared with the Coulomb and exchange matrix elements. However, after 
accelerating the two latter by, e.g., RI methods their evaluation becomes the most time consuming 
part in the Fock matrix construction. 

Considering the potential of the above-mentioned techniques to accelerate the Fock matrix
construction, certainly the fast calculation of ERIs is of prime importance. Based on the
current state of the art a rigorous screening of the integrals and density-fitting techniques are
the methods of choice. Since the integrals can be evaluated independently they are prone for
a parallelized evaluation, which will be discussed later. The exchange--correlation 
matrix elements require efficient numerical techniques for an accelerated calculation. 
Straightforwardly, the coarseness of the grids employed can be increased to decrease the 
computational effort within pre-defined accuracy bounds. 

\subsection{Density-matrix Construction}\label{sec:densitymatrix}

In conventional SCF-type electronic structure calculations the density matrix is calculated by a
full diagonalization of the Fock matrix. This approach has the advantage that the complete 
eigenspace can be obtained, i.e., including all virtual orbitals. A disadvantage is, however, that the scaling with 
respect to system size is cubic (and can be easily prohibitive in plane-wave calculations). Due to 
the small pre-factor of the diagonalization procedure for Gauss-type orbitals it mainly poses a 
problem for systems with several thousands of atoms. 

Almost all methods discussed here have been developed for problems where the
construction of the Fock matrix is very fast and therefore the construction of the density
matrix becomes the bottleneck of the calculations. An excellent review over methods, which avoid the
diagonalization in semi-empirical calculations can be found in Ref.~\citen{daniels1999}. An
assessment of density-matrix methods for self-consistent-field calculations by comparing 
purification and minimization methods has recently been published by Rudberg 
{\it et al}.\cite{rudberg2011}

The diagonalization of the Fock matrix does not directly yield the density matrix but the
coefficient matrix $\boldsymbol{C}$ which is then contracted to obtain the density matrix
$\boldsymbol{P}$. Although the density matrix is a local quantity due to the nearsightedness of the 
electrons\cite{kohn1996}, the coefficient matrix is not. Therefore, a great reduction of computation 
time would be achieved calculating the density matrix directly from of the Fock matrix. 

The so-called energy minimization techniques exploit the fact that the correct density matrix
minimizes the expression $\textnormal{tr}[\boldsymbol{P}\boldsymbol{F}]$. This minimization,
however, has to be done under the constraints that the density matrix fulfills the idempotency
condition and the trace condition, which requires that the trace of the density matrix yields the
number of electrons. By contrast, the diagonalization procedures explicitly fulfill the idempotency
condition. Li, Nunes, and Vanderbilt proposed the functional
\begin{align}
 E\left(\boldsymbol{P}\right) = \textnormal{tr}\left[\left(3 \boldsymbol{P}^2-2 
                                     \boldsymbol{P}^3\right)
 \left(\boldsymbol{F}-\mu \boldsymbol{I}\right)\right] \,\, ,
\end{align}
as a method to compute the density matrix by implicitly fulfilling the idempotency condition\cite{li1993}. 
Here, $\mu$ denotes the chemical potential and $\boldsymbol{I}$ the identity matrix.
Although this functional requires the density matrix to be set up in an orthogonal basis, there is 
no major problem to obtain a formulation for non-orthogonal basis sets, for which the energy 
functional is modified to include also the overlap matrix.\cite{nunes1994} Independently, Daw 
proposed a similar approach in 1993\cite{daw1993} that applied additionally steepest-decent 
iterations to minimize the functional. Alternatively, a conjugate-gradient approach can be chosen, 
which is then called the conjugate-gradient density-matrix search method\cite{millam1997,
challacombe1999}. Then, the Fock and the density matrix are transformed into an orthonormal basis. 
Another advantage of this approach is that the chemical potential does not need to be known in 
advance. Other techniques employ for example curvy steps\cite{helgaker2000} or include second 
derivatives\cite{ochsenfeld1997} to obtain faster convergence. All these methods employ a 
purification of the matrix proposed by McWeeny\cite{mcweeny1960} in order to fulfill the idempotency 
condition of the density matrix. The so-called sign matrix methods\cite{beylkin1999,vandevondele2012} 
follow a different way to achieve this by expressing the density matrix in terms of the sign matrix 
function, which can be computed by iterative schemes. Also possible is the introduction of a penalty 
functional for violating the idempotency condition.\cite{kohn1996} However, the penalty functional 
is difficult to employ since it cannot be minimized by standard methods. Linear scaling of these 
methods is obtained by weakening the idempotency condition through restricting the minimization to 
localized density matrices with density matrix elements corresponding to a distance larger than a 
certain threshold forced to be zero.  Other methods, which are not that common in {\it ab initio} 
electronic structure calculations, shall just be mentioned here for the sake of completeness. 
Examples are the Fermi Operator Expansion\cite{goedecker1995,goedecker1994} and the Fermi Operator 
Projection\cite{goedecker1995b} method.

Another possibility to minimize the energy without explicitly imposing the orthonormality condition, 
as it is done by full diagonalization, is to minimize with respect to orbitals only with an implicit 
orthonormalization constraint. The Orbital Minimization\cite{kim1995,mauri1994,mauri1993,
ordejon1998} or the Optimal Basis Density Matrix Minimization\cite{hernandez1995,hierse1994} methods 
are typical examples. However, they are mainly applied in large tight-binding or semi-empirical
calculations.\cite{daniels1999} A linear-scaling behavior can be achieved in these methods by 
carrying out the minimization with respect to localized orbitals\cite{galli1992} (for example, by 
searching only over functions which non-zero values inside a specified region). This approach would 
not introduce any approximation if the localized orbitals could be obtained by a unitary 
transformation of the occupied eigenstates of the Roothaan--Hall equation.

\subsection{Acceleration of SCF Convergence}

Assuming that the build-up of the Fock matrix and the subsequent calculation of the density matrix 
are fast enough for Real-time Quantum Chemistry, the SCF procedure  poses severe obstacles. 
The sheer number of SCF iterations strongly depends on the atomic 
configuration and is hardly predictable. Therefore, it cannot be guaranteed that the wave function 
and the gradients are available in time. From this point of view, for Real-time Quantum Chemistry it 
would be desirable to have a method which completely avoids any iterative methods. However, in all 
true {\it ab initio} electronic structure calculation an iterative procedure is unavoidable, since the 
Fock matrix depends on the density matrix elements. Consequently, to be able to obtain a real-time 
calculation, one has to focus on reducing the number of iterations to a minimum. Performing the structural manipulations 
in small steps is therefore the key for a working Real-time Quantum Chemistry implementation.

The steps discussed in the previous sections (Fock-matrix assembly and density-matrix 
calculation) are both part of a single step in the self-consistent-field procedure.
The convergence of SCF iterations strongly depends on the first guess for the molecular orbitals 
and on the nuclear configuration. In a reactivity study, however, it may happen that
configurations of the atoms occur, for which the SCF procedure converges only slowly or 
not at all. It is thus of utmost importance to have methods at hand, which yield 
stable and fast converging SCF iterations. 

Direct inversion in the iterative subspace (DIIS)\cite{pulay1980,pulay1982} is a widely employed 
technique to accelerate and stabilize SCF iterations. Error vectors from previous iterations are 
calculated and minimized in a least-squares sense. Accordingly, previous iterations are utilized to 
extrapolate the Fock matrix in the next iteration. Quite closely related to the DIIS method are 
techniques called Fock matrix dynamics\cite{pulay2004} or electron density extrapolation
\cite{niklasson2006} which are common in the field of BO molecular dynamics. Instead of accelerating 
the SCF algorithm itself the whole single-point SCF calculation is accelerated by 
extrapolating the information from previous time steps of the simulation. It is assumed that in 
between two steps the nuclear coordinates change only little so that the molecular orbitals, Fock 
matrix elements, or the density do not differe much from one step to the next and hence are a good starting point for the 
electronic structure optimization for the new nuclear configuration. Although SCF acceleration
schemes have a long history, significant improvements can still be made as highlighted by the
augmented Roothaan--Hall method\cite{host09}.

The pseudo-diagonalization technique\cite{stewart1982} is based on the observation that only for the
first and the last SCF iteration a full diagonalization is necessary. In between it is sufficient to
eliminate all Fock matrix elements connecting the occupied and virtual molecular orbitals by 
unitary transformations. As a consequence, the diagonalization of the Fock matrix blocks corresponding to the 
virtual--virtual and the occupied--occupied molecular orbitals can be avoided, which greatly reduces 
the time of each SCF iteration.

In cases of a too narrow gap between the highest occupied and the lowest unoccupied molecular
orbital, slow convergence or even divergent SCF iterations can occur. In such cases 
level shifting\cite{saunders1973} can be applied to enlarge the gap and therefore avoid a
mixing. With this procedure the problem of divergence, slow convergence, or oscillating behavior can 
be cured in many cases. 

In all techniques discussed above the construction of the density matrix and the optimization of
the molecular orbitals in the SCF iterations were independent processes. However, for methods based
on minimization of an energy functional for constructing the density matrix, methods have been 
developed that combine the density matrix optimization and the self-consistent-field iterations in 
one single optimization loop.\cite{car1985,stich1989,teter1989,payne1992} The idea was developed for 
the coupled electron-nuclei problem (Car--Parrinello molecular dynamics\cite{car1985}) and is
therefore often called the ``molecular dynamics'' method, but it can also be applied in situations were
the nuclei are kept fixed. The difference to conventional matrix diagonalization procedures to
solve for the eigenstates is that the variational principle is applied in a dynamic fashion and all
eigenstates are determined simultaneously. The dynamic variables are here the coefficients of
the basis functions with a fictitious electon mass. To exploit this methodology 
for a Real-time Quantum Chemistry implementation one could require that the manipulations are done in a 
continuous fashion, i.e., along a 'trajectory' with sufficiently small configuration-change steps.

\subsection{Gradient Calculation}

As outlined in the beginning of this section the calculation of the energy gradient with respect to 
the position of the nuclear coordinates involves contributions from each term in the electronic 
energy. 
An efficient force evaluation for large molecular systems in the framework of pure DFT has recently
been proposed by Reine {\it et al}.\cite{reine2010} by combining screening with a fast multipole method. 
In addition the calculation of the Coulomb contributions were accelerated by employing a density-fitting 
scheme\cite{versluis1988,fournier1989} with auxiliary basis sets. There are 
also efficient gradient implementations available that do not not employ density fitting\cite{burant1996,shao2001}.

\subsection{Subsystem Approaches}

To divide a molecular system under study into smaller subsystems is a key to the molecular-model 
approach discussed in the first two sections of this work and offers the possibility to reduce the 
computational effort further. As through a magnifying glass, the reactive part/region of a molecular assembly
can be embedded in a spectator background and this magnifying lens can even be moved around in the whole system\cite{gasc06}.
So-called combined quantum mechanics/molecular mechanics (QM/MM)\cite{senn2007,lin2007} approaches are widely employed to enable calculations of large enzymatic 
systems. In QM/MM the reactive part is treated quantum mechanically and the surrounding environment 
is modeled by classical force fields. By contrast, QM/QM methods apply the laws of quantum mechanics 
to all subsystem but may treat them with different methodologies\cite{humbel1996}. 

Depending on how the subsystems are embedded into each other a different level of acceleration can 
be achieved. Completely independent subsystem methods---also called Divide-and-Conquer (D\&C) 
approaches---facilitate a massively parallelized calculation. In the original formulation 
of the D\&C approach the independently calculated density matrices of the subsystems were merged to 
yield the full density matrix\cite{yang1991}. Although the computation of the density matrices is 
done independently, the surrounding of a fragment is accounted for by buffer regions around the 
fragment. This approach does not only accelerate the calculation of the density matrix but also the 
calculation of the Fock matrices in the SCF calculations, which is the reason why this methodology is treated 
in this separate section and not together with other density matrix construction schemes. One 
difficulty for Real-time Quantum Chemistry is, however, that the calculation of the D\&C force for 
the full system is not well defined. But there are cures available\cite{zhao1995}. The partitioning 
can also be carried out for other quantities than the density-matrix. For instance, in the 
fragment molecular orbital theory the fragmentation is done at the level of the molecular 
orbitals.\cite{fedorov2007} For a recent and comprehensive review on fragmentation methods we refer 
to Ref.~\citen{gordon2012}

Subsystem techniques  allow for an in principle exact embedding of the subsystems thus 
retaining only the approximations introduced by choosing different electronic-structure methods for the subsystems. In 
the framework of density functional theory such methods have been proposed and are widely employed, 
for instance, to account for solvent effects\cite{wesolowski1993,wesolowski2006,iann06,fux-08,kiew08,jacob2008fde,neugebauer2010,gomes2012}.

To assess the potential acceleration gained by any subsystem approach, one has to consider two 
aspects: (i) how small can the subsystems be chosen and (ii) how are they connected to each 
other, if at all. This affects not only the computation but also the accuracy. But since we are
interested in local phenomena, the fragmentation is 
often already implied by the structure of the molecular system itself. Clearly, the smaller the
subsystems can be chosen and the less one has to account for embedding effects, the better for 
Real-time Quantum Chemistry.

From the Real-time Quantum Chemistry point of view, Divide-and-Conquer and density embedding 
approaches are appealing, since they allow the greatest reduction in computation time if the whole 
system needs to be treated quantum mechanically. If, however, large parts of the molecular system 
are not directly involved in a reaction, but rather serve as a dielectric 
environment, then QM/MM methods are most suitable.

\subsection{Technical Aspects: Parallelization and Special Hardware}\label{sec:spechardware}

Essentially all of the most popular quantum chemistry codes can be run in parallel. However, these software approaches 
usually have a rather high overhead making them not the best candidates for Real-time 
Quantum Chemistry applications. But
electronic structure calculations have not only been accelerated by developing increasingly
efficient algorithms but also by employing latest and even specialized hardware. 

Although graphical processing units (GPUs) were originally designed for the fast rendering of three dimensional 
graphics, they can be employed for the acceleration of quantum chemical calculations. In the past 
years the development and application of special algorithms for  quantum chemical calculations 
that exploit the computing power of GPUs has gained considerable attention.\cite{yasuda2008,
yasuda2008b,ufimtsev2008,ufimtsev2009,ufimtsev2009_2,luehr2011} This trend is due to a significant 
effort undertaken to program quantum chemical algorithms specifically designed for GPUs but also due 
to new graphic cards produced with a focus on scientific calculations. For example, the widely used 
GAMESS US package\cite{schmidt1993} or the Terachem program\cite{ufimtsev2009} are able to 
efficiently exploit the advantages of GPUs for electronic structure calculations. Compared to 
calculations performed only on the central processing unit (CPU) of a computer they are able to 
achieve considerable speed-ups in the order of one magnitude.\cite{stone2010} Also in the field of 
semi-empirical calculations GPUs have attracted some attention.\cite{wu2012}

Massive parallelization is also possible on processor architectures other than GPUs. For 
Kohn--Sham DFT calculation, for instance, the application of processors from ClearSpeed Technology 
Ltd.\ has been reported to accelerate electronic structure and gradient calculations. 
\cite{brown2008,brown2010} Besides these developments for specific processor types, there are also 
some more general considerations about how to exploit the advantages of emerging new processor types
available in the literature.\cite{ramdas2008,ramdas2009} A comprehensive overview of special
processors and their potential for electronic structure calculation can be found in 
Ref.~\citen{ramdas2008b}. In the field of classical molecular dynamics simulations the so-called 
ANTON processor developed by Shaw {\it et al}.\cite{shaw2007} is a successful attempt to build such 
a special purpose processor. Also for electronic structure calculations special processors have 
been designed; namely ERIC, the ERI Calculation specific chip-multiprocessor\cite{nakamura2005} or 
the molecular orbital calculation specific embedded high performance computing (EHPC) 
system\cite{umeda2009}.

Almost all special processor types mentioned here accelerate the electronic structure calculations
by parallelization of the ERI calculation. This is the reason for the success of GPUs but also of 
high performance clusters like, for instance, the IBM Blue Gene series\cite{bluegenep2008,
fletcher2012}. The calculations of ERIs is in almost all quantum chemical calculation the 
bottleneck because of their sheer number. 
Specialized hardware which allows for a massive parallelization can directly accelerate the 
calculations and not only improves the scaling behavior. Thus, exploiting these techniques is 
certainly imperative for Real-time Quantum Chemistry.

%

\section{Direct Haptic Quantum Chemistry}\label{sec:dhqc}

After having elaborated on the available and future quantum chemical methods for Real-time Quantum
Chemistry implementations, we shall now discuss their benefits for Haptic Quantum Chemistry\cite{marti2009,haag2011} 
and subsequently discuss their capabilities in an out-of-the-box application presented in the next section.

\begin{figure}[h]
\begin{center}
 \includegraphics[width=\textwidth]{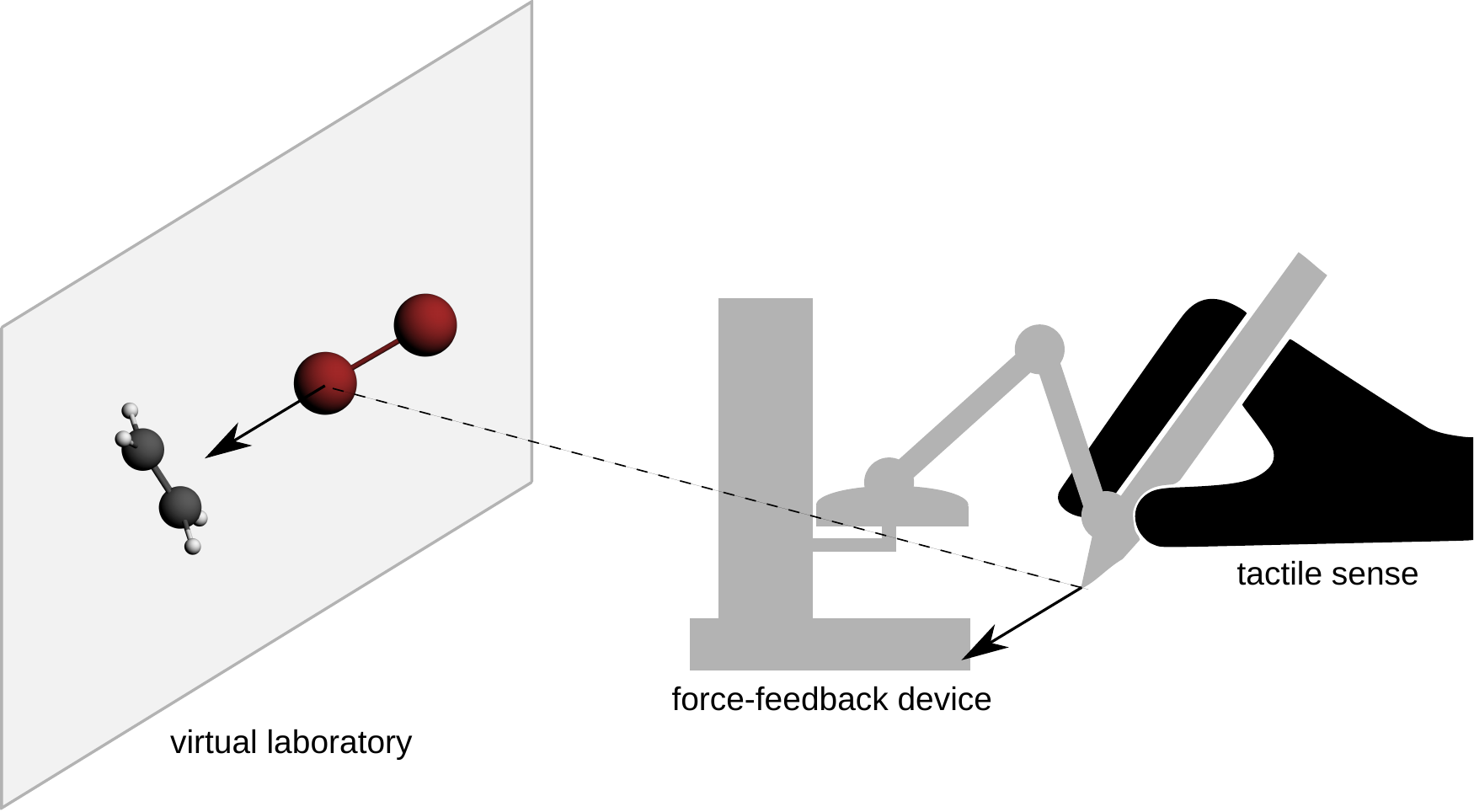}
\end{center}
\caption{The haptic device in Haptic Quantum Chemistry showing how a bromine molecule is
moved towards an ethene molecule. The tip of the haptic pointer corresponds to the bromine atom
next to the ethene.\label{fig:hapticdevice}}
\end{figure}

The concept and implementation of Haptic Quantum Chemistry as presented in Refs.~\citen{marti2009,
haag2011} is to employ a force-feedback device as depicted in Fig.~\ref{fig:hapticdevice} as an 
input and an output tool allowing for an intuitive manipulation of molecular structures while feeling 
the gradients on the manipulated atoms as forces. Such approaches are referred to 
as haptic enabled interactive molecular visualizations systems in the literature.\cite{marchese2009} There are a few 
such methods already available, but they only employ classical force fields to calculate the forces 
rendered by the device.\cite{bayazit2001,laiyuen2005,bidmon2007,wollacott2007} Thus, they prohibit 
the study of chemical reactions, which would require the ability tp form and break chemical bonds.

In our current set-up, the haptic device is a pen-like pointer which allows the user to manipulate 
objects in a virtual reality framework and, at the same time, to physically experience a force feedback (cf.\ 
Fig.~\ref{fig:hapticdevice}). 
Hence, one is able to feel the curvature of the potential energy surface of the manipulated nuclear coordinates 
in the whole system. The visual presentation of the structure, the gradients on the 
atoms not manipulated with the device, the orbitals,  the electron density and other properties
allow one to perceive complex information in a very intuitive 
way compared with the sole visual presentation. Hence, the haptic quantum chemical approach immerses 
the scientist more into the scientific problem. Even deeper immersion can of course be achieved by
employing 3D displays or 3D glasses and other techniques from the field of virtual reality.

Since the human haptic sense is much more sensitive than the visual sense, haptic devices usually
have an update rate of about $1-4\,\textnormal{kHz}$. By contrast, to create the illusion of
a smooth movement for the visual sense only $25\,\textnormal{Hz}$ are necessary. To circumvent
this problem, in  Haptic Quantum Chemistry\cite{marti2009,haag2011} so far the forces are not 
calculated directly but are obtained from interpolating single-point gradients $\{\boldsymbol{g}\}$. 
The force $\boldsymbol{f}_I$ acting on an atom $I$ is calculated from the interpolated gradient 
$\boldsymbol{\widetilde{g}}$ by 
\begin{align}
 \boldsymbol{f}_I &= -\,\boldsymbol{\widetilde{g}}_I \,\,,
\end{align}
where the gradient of the energy is given by
\begin{align}
 \boldsymbol{g}_I &= \boldsymbol{\nabla}_I \, E_\textnormal{BO} \,\,.
\end{align}
Here, $\boldsymbol{\nabla}_I$ denotes the three Cartesian derivatives with respect the nuclear 
coordinate of atom $I$. The computationally inexpensive interpolation facilitates the very fast 
calculation of the forces necessary for the high update rate of the haptic device. The main drawback 
is, however, that a coarse-grained gradient field of the molecular system needs to be calculated in 
advance (though it can be refined during haptic exploration as more data points can be calculated in the 
background).

The pre-calculation of single-point gradients becomes, however, more and more demanding if parts of
the molecular structure are allowed to relax. The molecular structure at one specific position in
the configuration space of the mobile part is no longer unique and depends on the trajectory leading
to it. As a consequence, one has to design algorithms that keep track of the history of the actual
haptic exploration run. However, employing a Real-time Quantum Chemistry framework would allow us to 
circumvent this problem as the quantum chemical data is always immediately available so that no 
history (trajectory) needs to be stored. We may call this approach Direct Haptic Quantum Chemistry 
(D-HQC).

\begin{figure}[htb]
\begin{center}
 \includegraphics[width=\textwidth]{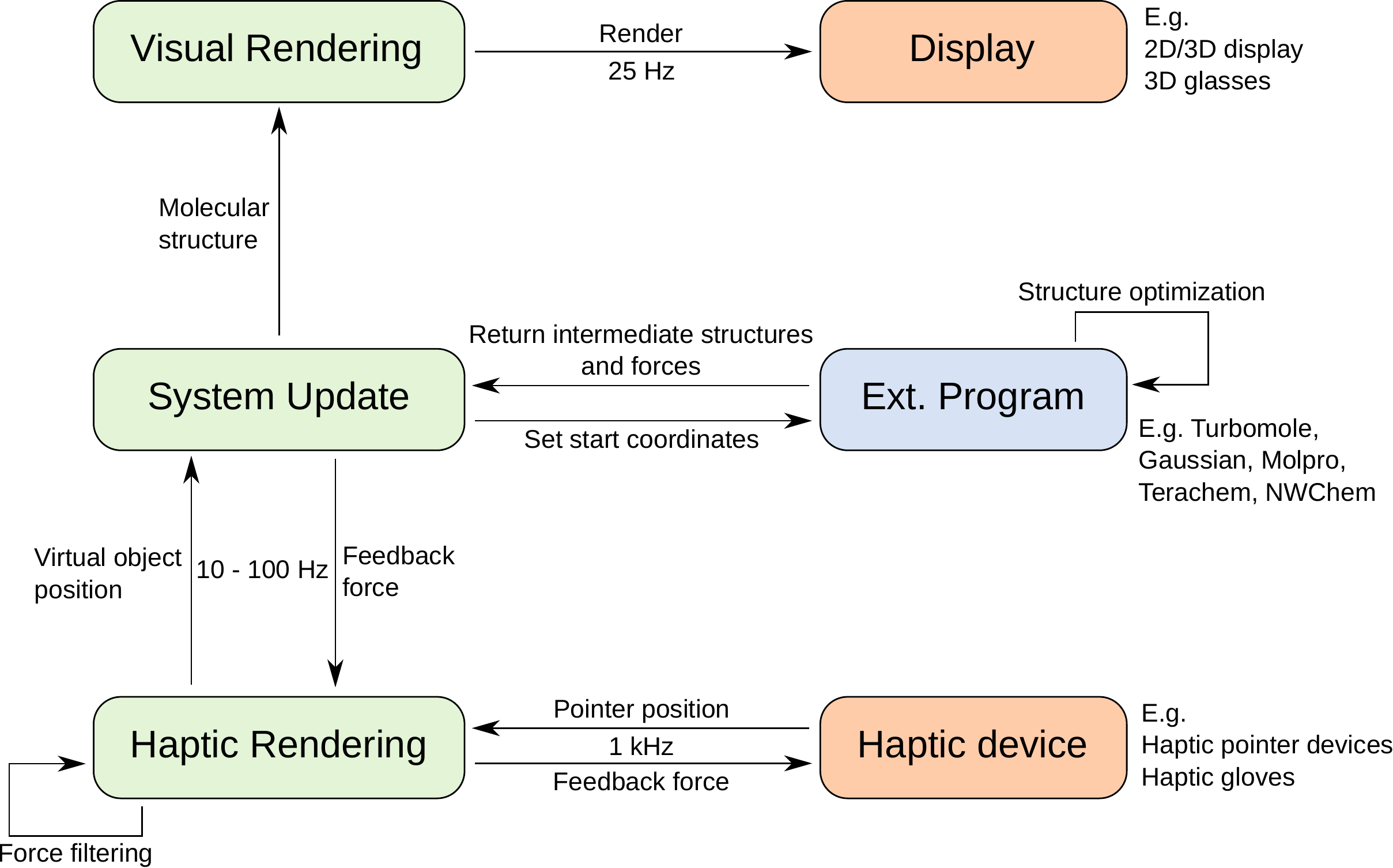}
\end{center}
\caption{Program structure of a Direct Haptic Quantum Chemistry
implementation.\label{fig:progstruct}}
\end{figure}

At first glance, the high update rate (more than 1 kHz) of the haptic device requires an update 
rate of a millisecond if the gradients from the electronic structure calculation 
were directly rendered by the device. But as it has been shown in the field of interactive fluid 
dynamics simulations, the servo loop of the haptic device and the simulation loop can be separated 
and can operate with different update rates.\cite{baxter2004} Both loops are connected by a shared 
memory from where the servo loop constantly reads the current force written by the simulation loop 
as soon as the new force is available. To reduce the occurring artifacts the stepwise force 
update is smoothed by a force filter. With this technique the update rate of the molecular system 
can be lowered below $100\textnormal{Hz}$. Accordingly, D-HQC would require that within a few hundred 
millisecond a new gradient on the manipulated atoms has to be available 
and the other atom positions have to be relaxed 
(Fig.~\ref{fig:progstruct}).

This almost instantaneous relaxation of the whole structure is, however, not always wanted. By 
contrast, it might be even desired to be able to alter the molecular structure faster than the 
system relaxes in order to simulate non-adiabatic changes. Therefore, the relaxation does not need 
to be instantaneous as long as the force change due to the relaxation can be rendered smoothly. For 
almost adiabatic changes the structure alterations have to be slowed down or done in small steps so 
that the structure of the whole system can relax. Altering the structure only in very small 
increments also speeds up the electronic structure optimizations, since the molecular orbitals 
change only very little. The program could force the user to perform only small changes (slow movements) by applying re-stalling
forces on the haptic device. 

D-HQC can be described as probing the potential energy surface in the configuration subspace spanned by the manipulated atoms of the
molecular system. For a more formal description of the force calculation in D-HQC the set of nuclei 
$I$ is partitioned into nuclei controlled by the haptic device $I^{\prime}$ and the remaining nuclei
$I^{\prime\prime}$. The force on a nucleus $I^{\prime}$ is then calculated as the negative spatial 
derivative of the total energy, which is minimized with respect to the basis set $\{\phi_i\}$ and 
the nuclear coordinates of the remaining nuclei
$\left\{\boldsymbol{R_{I^{\prime\prime}}}\right\}$.
\begin{align}\label{eq:directhqcgradient}
 \boldsymbol{f}_{I^\prime} = - \boldsymbol{\nabla}_{I^\prime} \underset{\{\phi_i\},
 \{\boldsymbol{R}_{I^{\prime\prime}}\}}  {\textnormal{min}} E_\textnormal{tot} \left[ \left\{ \phi_i
 \right\}; \left\{ \boldsymbol{R}_{I^\prime1} \boldsymbol{R}_{I^{\prime\prime}} \right\} \right]
\end{align}
with the total energy functional $E_\textnormal{tot}$ is given by
\begin{align}
 E_\textnormal{tot}\left[ \left\{ \phi_i \right\}; \left\{ \boldsymbol{R}_{I^\prime},
 \boldsymbol{R}_{I^{\prime\prime}} \right\}\right] = \left\langle \Psi_0(\left\{
 \boldsymbol{R}_{I^\prime},
 \boldsymbol{R}_{I^{\prime\prime}} \right\}) \left|
 \hat{H}_\textnormal{el}^\textnormal{eff} \right| \Psi_0(\left\{ \boldsymbol{R}_{I^\prime},
 \boldsymbol{R}_{I^{\prime\prime}} \right\}) \right\rangle \,\, .
\end{align}

\section{Examples}\label{sec:testcalcs}

Many of the above-mentioned developments of linear and sub-linear scaling methods are available
in standard program packages, though the ultimate package for real time quantum chemistry has
not been developed yet---mostly for the reason that each package has been designed to serve a
certain purpose. For example, huge molecular models have already been studied with the {\sc CP2K}
\cite{lippert1997,lippert1999,vandevondele2005} program that employs mixed Gaussian and plane wave 
basis sets in AIMD simulations. Here, we choose the very efficient DFT modules of the 
{\sc Turbomole} package\cite{treutler1995,vonarnim1998} combined with our D-HQC setup to demonstrate 
that such studies are in reach for molecular models of relevant size. The calculations exploit 
density fitting, effective core potentials and small basis sets. It is clear that the resulting
accuracy then does not necessarily live up to the current standard. However, this is also not 
decisive as a reaction pathway recorded during a D-HQC exploration can always be relaxed on a more
accurate potential energy surface.

\subsection{D-HQC for a bromine molecule approaching an ethene}

\begin{figure}[H]
\begin{center}
 \includegraphics[width=0.7\textwidth]{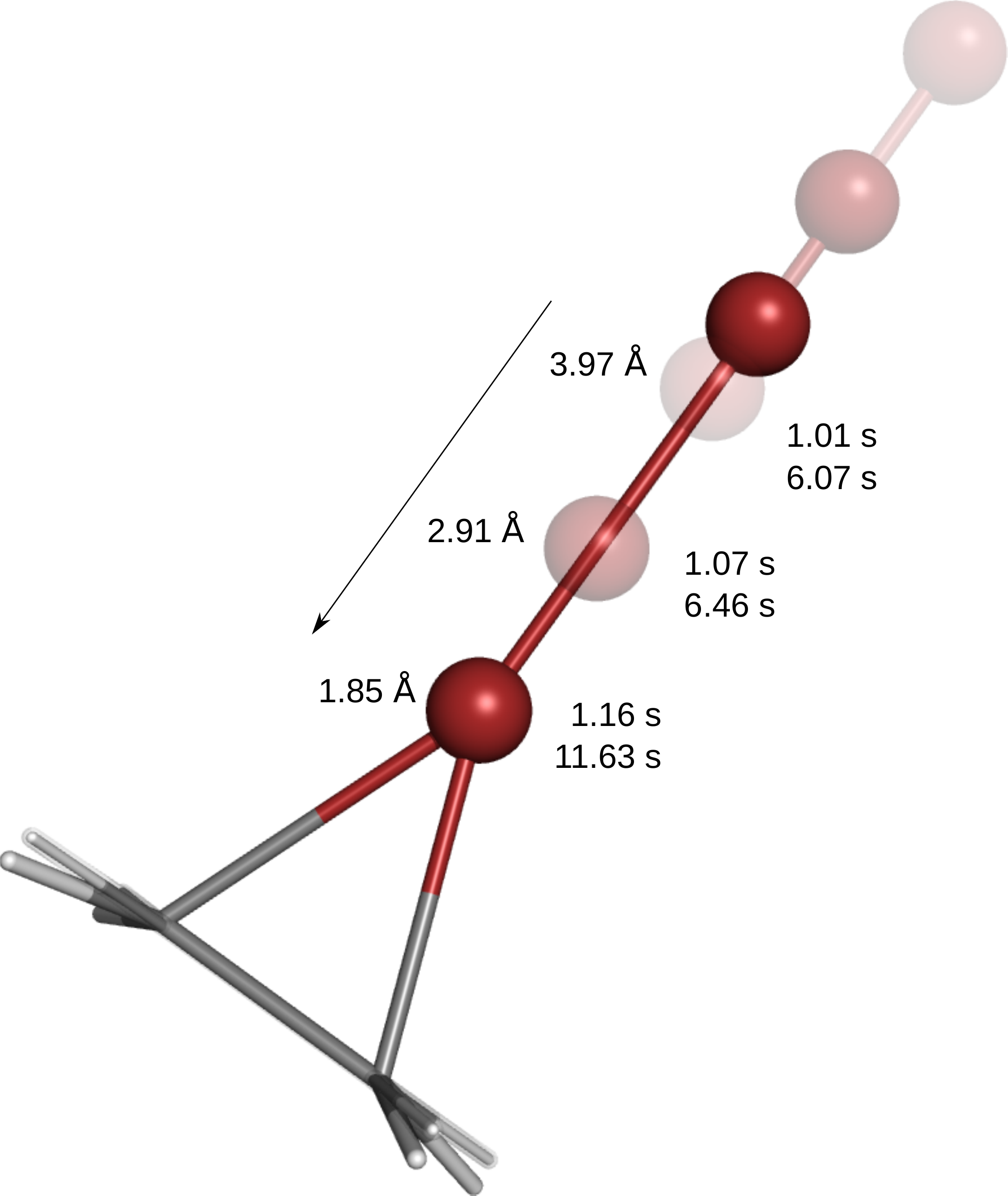}
\end{center}\label{fig:bromethen}
\caption{Exploratory calculation for the reaction of a bromine molecule (in red) with ethene. The distance from the
closest bromine atom to the ethene is written above this bromine atom. Below it, the average time per single
point in the optimization and the overall time until structure convergence are given.}
\end{figure}

As depicted in Fig.~\ref{fig:bromethen} in this exemplary study a bromine molecule was pushed onto an
ethene molecule to probe its reactivity. In this setup the position of the bromine atom closest to
the ethene was dragged towards the ethene molecule so that the relaxation at each step had the
bromine to ethene distance as a constraint. In this way a trajectory in the subspace spanned by the
position of the bromine atom was generated. The distance was incrementally decreased by 0.5 Bohr.
The resulting distances in {\AA}ngstroms are given in Fig.~\ref{fig:bromethen}, which shows three exemplary points
from the trajectory. In addition, the average time per structure relaxation step, i.e., the
time for updating the system's structure, 
as well as the overall time needed to converge the structure (second line)  
are depicted.

Following the results of the discussion about basis sets we chose for the 
bromine atoms the Stuttgart ECP-28-MWP pseudo potential\cite{schwerdtfeger1989} and the def2-SV(P) 
basis set, for the carbon atom the def2-SV(P) basis set\cite{schaefer1992} and for the hydrogen atoms the 
STO-3G HONDO basis set\cite{hehre1970,hehre1969} in order to obtain a very fast calculation of single-point 
energies and gradients. The calculations were performed employing the BP86 
exchange--correlation density functional\cite{dirac1929,slater1951,vosko1980,becke1988,perdew1986} on a coarse
numerical grid. In addition, also the resolution-of-identity technique (RI)\cite{eichkorn1995} 
was applied to accelerate the calculations. 

The execution times in Fig.~\ref{fig:bromethen} show that the time needed to update the system is
almost constant during the trajectory, but the structure optimization time increases when the reactants get closer, which
indicates that it needs more steps to converge. As it was outlined before, the update rate is the 
important quantity for a real-time experience. The execution times per update step can be
shortened by sampling the trajectory in smaller steps, which means that the SCF procedure can
converge faster since the wave function does not change too much from step to step.  

The computations here were performed by running the 
individual {\sc Turbomole} modules sequentially. Note that there is still room for improving the efficiency in terms of 
passing the information from one calculation to the next and avoid read/write accesses to the hard 
disk as these processes have not yet been optimized for a D-HQC implementation in standard programs
like {\sc Turbomole}.

\subsection{D-HQC for an S$_\textnormal{N}$2 reaction of fluoride with chloromethane}

Another example with a more pronounced effect of structural relaxation of the remaining atom positions is 
expected to demonstrate how this influences the systems update rate: a S$_\textnormal{N}$2 type 
reaction. In this example a fluoride ion approaches a chloromethane molecule and replaces 
the chlorine atom. The trajectory was recorded by moving the fluoride anion in steps of 0.1 Bohr 
towards the C atom. At each step the structure of the remaining atoms was optimized.
The intermediate electronic structure optimizations were carried out again with small basis sets 
(def2-SVP \cite{schaefer1992} for the C, F, and Cl atoms and STO-3G HONDO \cite{hehre1970,hehre1969} 
for the H atoms) in combination with effective core potentials (ECP-10-MWB\cite{schwerdtfeger1989} 
for Cl and ECP-2-SDF for F) and BP86/RI on a coarse numerical grid.
To reduce the number of SCF cycles in each geometry optimization cycle the molecular orbitals from the preceding point of the 
trajectory were taken. 

\begin{figure}[H]
\begin{center}
 \includegraphics[width=\textwidth]{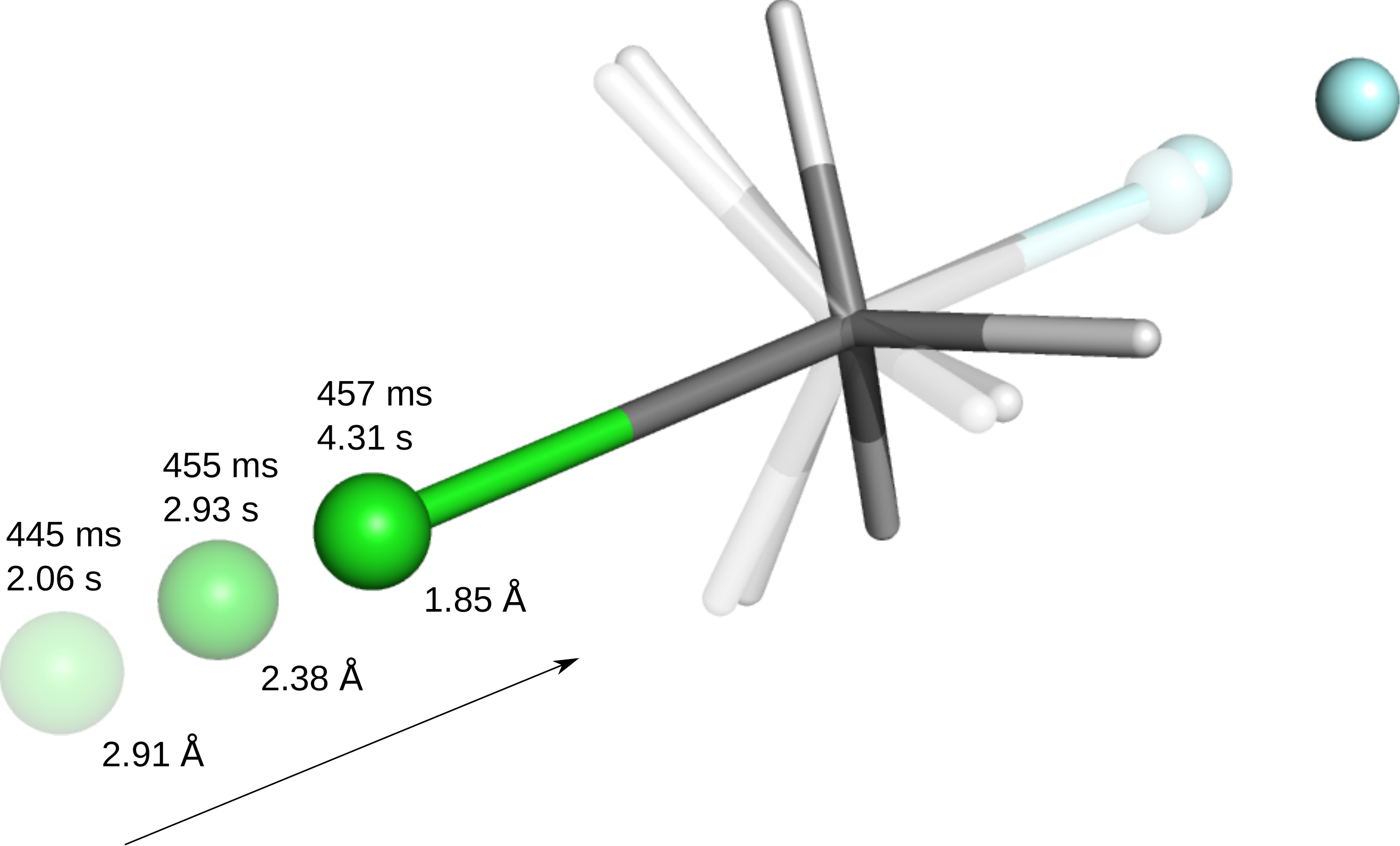}
\end{center}\label{fig:sn2}
\caption{D-HQC exploration for the S$_\textnormal{N}$2 reaction of fluoride (green) and
chloromethane. The chlorine atom is printed in blue. The distance from the attacking fluorine anion 
to the central carbon atom is written below the fluorine atoms. Above the atoms the average time per 
single point in the optimization (in ms) and the overall time until structure convergence (in s) 
is given.}
\end{figure}

In Figure \ref{fig:sn2} three intermediate points of the trajectory are shown. The average time
per electronic structure calculation in the structure optimization is almost constant and on the
order of several hundred milliseconds. The overall time for the structure optimizer to reach
convergence is, however, not constant and increases significantly when the attacking fluoride
approaches its position in the transition state. The detailed timings show again that not the 
electronic structure optimization but the increased number of cycles in the structure optimization 
give rise to the increased overall time. As already discussed in the previous section this is not a 
severe issue as only the system updates have to be fast, which is this case in this example. The 
user would have to slow down the movement of the fluoride when approaching the C atom in order to 
obtain a reasonable minimum energy path. For a non-adiabatic simulation the movement can be faster although the 
remaining atoms of the system are not able to relax in time.

\section{Conclusions}\label{sec:conclusion}

The possibility and necessity of obtaining the result of quantum chemical calculations in real time 
is beneficial in many respects for studying the reactivity of chemical systems and may change the way 
how quantum chemistry is done in the future. Not only the ever increasing amount of information 
provided by calculations but also the inherent complexity of chemical problems calls for new 
approaches like (Direct) Haptic Quantum Chemistry that require a Real-time Quantum Chemistry 
framework. The overview of currently available techniques to accelerate calculations provided here
clearly showed that Real-time Quantum Chemistry is in reach and will be possible for relevant system 
sizes in the near future. 

The evaluation of existing algorithms and technology for Real-time Quantum Chemistry also 
demonstrated, however, that a paradigm change is needed. Almost all techniques presented here 
were not specifically designed to allow quantum chemical calculations of energies and gradients in 
real time. The aim of their development was the overall scaling behavior to allow the treatment of 
ever larger molecules or molecular systems. For Real-time Quantum Chemistry the focus needs to be on 
reducing the actual execution time for a fixed system size to around $100\,\textnormal{ms}$. Although 
most of the currently available program packages in quantum chemistry have not been developed to 
allow the ultra-fast calculation of molecular systems consisting of $100 - 200$ atoms, the greatest 
potential for achieving considerable speed-up towards real time lies most probably in the activation 
of specialized hardware. Already in reach are calculations on GPUs which show a promising potential, 
but also completely new specialized hardware is desirable for Real-time Quantum Chemistry.

The overwhelming amount and the complexity of the data generated by current quantum chemical
calculations already limits their fast and intuitive evaluation. Haptic Quantum Chemistry offers a 
new approach to tackle this problem. The instantaneous availability of the wave functions and the 
gradients offered by Real-time Quantum Chemistry allows an even more convenient way of studying 
chemical reactivity, as we have discussed for the Direct Haptic 
Quantum Chemistry variant. The exploitation of the human haptic sense to present scientific data more 
intuitively is only a first step. A deeper immersion by employing techniques already 
developed in the field of virtual reality would be the ultimate goal of any development in this 
direction.

\section{Acknowledgments}
We gratefully acknowledge financial support through TH grant ETH-08 11-2.

\thispagestyle{empty}
\providecommand{\refin}[1]{\\ \textbf{Referenced in:} #1}

\end{document}